\documentclass[pre,reprint,twocolumn,superscriptaddress,floatfix,aps]{revtex4-2}
\usepackage{graphicx}
\usepackage{dcolumn}
\usepackage{bm}
\usepackage{xcolor}
\usepackage{float}
\usepackage{mathrsfs}
\usepackage{soul}
\usepackage{amsmath}
\usepackage{hyperref}
\usepackage{amsfonts}

\begin{document}


\title{Effect of higher-order interactions on tipping cascades on complex networks}
\author{Richita Ghosh}
\author{Manish Dev Shrimali}
\affiliation{Department of Physics, Central University of Rajasthan, Rajasthan, Ajmer 305 817, India}





\begin{abstract}
Tipping points are critical thresholds of parameters where tiny perturbations can lead to abrupt and large qualitative changes in the systems. Many real-world systems that exhibit tipping behavior can be represented as networks of interacting multistable units, such as vegetation patches or infrastructure networks, undergoing both pairwise and higher-order interactions. In this article, we explore how higher-order interactions shape the dynamics of tipping cascades in a conceptual system with tipping points. Numerical simulations on random, scale-free, and small-world networks reveal that higher-order interactions can induce cascades even at coupling strengths, where only pairwise interactions fail to do so. We also investigate the interplay of the pairwise and higher-order coupling strengths in random networks and illustrate the route to cascades through bifurcation diagrams.  These results have also been demonstrated on real-world social networks. Apart from this, we show that repulsive higher-order interactions suppress tipping cascades at coupling strengths where pairwise interactions would cause them, and shift the cascade route from a saddle-node to a supercritical pitchfork bifurcation. Our results highlight the critical role of higher-order interactions in shaping cascade dynamics, offering insights for anticipating and mitigating critical transitions in ecosystems, climate systems, and socio-technical infrastructures.
\end{abstract}
\maketitle


\section{Introduction}
In recent years, the study of tipping points has become a major point of interest in complex systems research \cite{gladwell2006tipping,milkoreit2018defining,van2016you,lenton2008tipping}. Tipping points or regime shifts are qualitative shifts in a system's state when a small perturbation pushes it past certain thresholds. Examples of tipping points abound in climate science, where the potential transition of climate tipping elements to a ``hothouse" state may signify serious risks to humanity \cite{ritchie2021overshooting,lenton2019climate}. Tipping points also appear in other complex systems, such as financial market crashes \cite{may2008ecology,kostanjvcar2016estimating}, shallow lakes\cite{scheffer2007shallow,hilt2011abrupt}, and other ecosystems \cite{scheffer2001catastrophic}. Thus, understanding the dynamics of tipping points is of paramount importance. 
\par In complex networks, tipping points represent critical thresholds of individual tipping elements, which can be regarded as subsystems of a larger system. Tipping elements occur in various systems, such as the Earth's climate system \cite{lenton2008tipping}, ecosystems \cite{duarte2012tipping}, and others. These elements may or may not be isolated and often interact across time and space \cite{rocha2018cascading,brummitt2015coupled}. Some examples of interacting tipping elements may be connected lakes \cite{van2017regime,scheffer2007shallow} or the climate tipping elements in the Earth system \cite{kriegler2009imprecise}. In such scenarios, if one tipping element crosses its tipping point, it becomes more likely that the other element will tip too \cite{klose2020emergence}. This is known as a tipping cascade since the tipping in the other element occurs solely due to the coupling between the elements. Such situations are becoming increasingly relevant in the modern world in the context of the impact on human societies due to climate tipping elements. \par Interactions between tipping elements may also occur through the mathematical framework of complex networks \cite{boccaletti2006complex,strogatz2001exploring,newman2003structure}. Networks are an indispensable tool for the study of complex systems. Complex networks consist of structures called nodes, and the links between these nodes are referred to as edges. In recent years, the physics of complex networks has been studied extensively to model coupled oscillators in power grids \cite{pagani2013power,arianos2009power}, food webs \cite{williams2002two}, transportation systems \cite{lin2013complex,zanin2013modelling}, or the collaboration of scientists \cite{ramasco2004self}. Tipping cascades have been explored in paradigmatic networks \cite{kronke2020dynamics} with pairwise interactions and also in models of opinion formation \cite{abraham1991computational}. 
\par But, all interactions cannot be comprehended through the mathematical framework of pairwise links, i.e., through the mathematics of one link connected to two nodes. Some interactions are better explained as group interactions. An example of such interactions can be a group of people engaging in conversation. In such scenarios, interactions cannot be broken down to pairwise links, and the mathematics of higher-order interactions (HOIs) is required. \par Higher-order interactions or group interactions can be encoded mathematically using the topology of simplicial complexes. A simplicial complex is a topological structure consisting of simplices of different dimensions \cite{battiston2020networks,battiston2021physics,majhi2022dynamics}. A simplex is written as a set of nodes $\mathrm{D} \in [x_0, x_1, \ldots, x_N]$ where the individual elements of the set denote each node of the simplex \cite{boccaletti2023structure}. A one-dimensional simplex is a link between two nodes, while a two-dimensional one refers to a triangle that describes the interaction between three nodes. A plethora of collective phenomena have been observed on complex networks with higher-order interactions, such as transitions to synchronization \cite{skardal2021higher,parastesh2022synchronization}, oscillation death \cite{ghosh2023first}, chimera states \cite{ghosh2024chimeric,kundu2022higher}, and others. Higher-order interactions can be attractive or repulsive in nature. Attractive and repulsive interactions have been explored from the pairwise point of view \cite{dixit2020aging,dixit2020static}. In complex systems with attractive or repulsive higher-order interactions, tipping may arise in social networks, such as during the spread of information or rumors, or in ecological networks, where the repulsive third species disrupts an attractive interaction, and shifts the ecosystem toward an alternative state.
\par In this article, we explore the dynamics of tipping cascades in networks with both pairwise and higher-order interactions. Each node or tipping element is modeled by the differential equation based on the normal form of the cusp catastrophe. This equation exhibits two saddle-node bifurcations and hysteresis. There are numerous examples of environmental tipping points whose conceptual models exhibit saddle-node bifurcations, e.g., the thermohaline circulation \cite{wunsch2002thermohaline}, the Greenland ice sheet \cite{levermann2016simple}, tropical rainforests \cite{staal2018resilience} adoption of new technology where people suddenly change to new platforms \cite{comin2004cross}, opinion formation \cite{abraham1991computational} among others. We show that higher-order interactions hasten tipping cascades, than when only pairwise interactions are active. That is to say, tipping cascades occur at lower coupling strengths with higher-order interactions than when only pairwise interactions are applied. We demonstrate this phenomenon with the help of random, small-world, and scale-free networks. The interplay of the higher-order coupling strength along with the pairwise one is also studied, and it shows how higher-order interactions influence the occurrence of such cascades. The bifurcation structures of the globally-coupled three-node system are also studied in detail, and we apply our model to real-world social networks and observe how higher-order interactions promote the occurrence of tipping cascades. Finally, we show how repulsive higher-order interactions can act as a critical tool in mitigating tipping cascades. The bifurcations are studied in detail, while it is observed that repulsive higher-order interactions mitigate tipping cascades on real-world networks. Thus, taken together, these findings highlight how both attractive and repulsive higher-order interactions govern the occurrence of tipping cascades. 

\section{Model}
\begin{figure}
    \centering
    \includegraphics[width=0.8\linewidth]{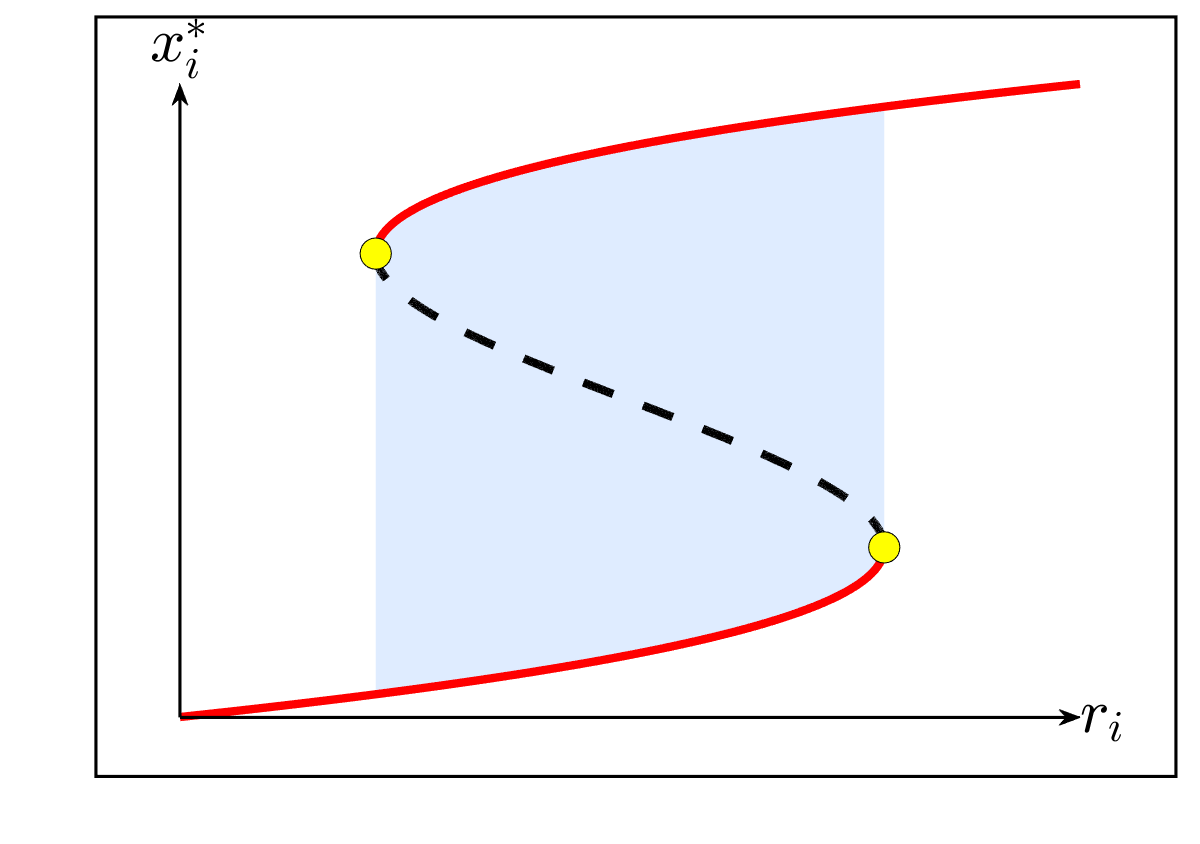}
    \caption{The bifurcation structure of a tipping element modeled by a cusp bifurcation is illustrated. The red lines show the stable steady states joined by the unstable state in dashed black. The yellow circles display the saddle-node bifurcations, while the shaded blue region shows the hysteresis area. }
    \label{fig:1}
\end{figure}

Each tipping element is modeled using the normal form of the cusp bifurcation. This bifurcation is used to signify nonlinear transitions between two alternate stable states. The normal form of the cusp bifurcation is written as:
\begin{align}
    \dot{x} &= -a(x-x_0)^3+b(x-x_0)+r 
    \label{Eq.1}
\end{align}
Here, the control parameters are $r$, while $a$ and $b$ are greater than zero. These parameters tune the strength of the system while $x_0$ represents the shift on the x-axis. The bifurcation diagram of Eq.~\ref{Eq.1} with respect to $r$ contains two stable branches, one upper and one lower, as depicted in red in Fig.~\ref{fig:1} along with a bistable or hysteretic region (blue shaded area in Fig.~\ref{fig:1}). The unstable branch is shown by the black dashed line. When the value of $r_i$ is low, the system stays on the lower stable branch. As the value of $r_i$ is slowly increased, at a critical value of $r$, the system transitions from the lower stable branch to the upper one. Now, if the value of $r_i$ is decreased, the state of the system stays on the upper branch, and at another critical value of $r_i$, the system drops down to the lower stable branch. Eq.~\ref{Eq.1} serves as a minimal model for multiple systems with alternative stable states and hysteretic properties, such as ecosystems \cite{kefi2016can}, thermohaline circulation \cite{stommel1961thermohaline}, and ice sheets \cite{garbe2020hysteresis}. 
\par Next, we consider a network of coupled tipping elements undergoing both pairwise and higher-order interactions. The model equation is written as: 
\begin{widetext}
\begin{align}
    \frac{dx_i}{dt} &= -a(x_i-x_0)^3 + b (x_i-x_0)+r_i + \epsilon_1\sum_{j=1,j\neq i}^{N} A_{ij}x_j + \epsilon_2\sum_{j=1, j\neq i}^{N} \sum_{k=1,k\neq i}^{N} B_{ijk}x_jx_k 
    \label{eq:model}
\end{align}  
\end{widetext}
Here, $\epsilon_1$ and $\epsilon_2$ are the coupling strengths, while $A$ and $B$ are the adjacency matrices of the pairwise and higher-order interactions, respectively. $A_{ij}=1$ if there is a link between the $i^{th}$ tipping element to $j^{th}$ tipping element and $0$ otherwise. The element of the tensor $B$, denoting the higher-order interactions, is given as $B_{ijk}=A_{ij}A_{jk}A_{ki}$. If there's a three-way interaction between the tipping elements $i$, $j$, and $k$, $B_{ijk}=1$, and otherwise $B_{ijk}=0$. Eq.~\ref{eq:model} has been simulated using the \textit{odeint} function from the SCIPY python package. The initial conditions for all the elements have been taken as zeros. The control parameter $r$ of the first tipping element is set to $0.2$ for all of the cases considered in this article. The rest of the elements of the vector $r_i$ are set to $0$, i.e., the $r$ of all the tipping elements except the first one are zero. The parameters $a$ and $b$ are considered to be $4$ and $1$ respectively, while $x_0=0.5$. For this set of parameters, two stable states are present in the system at $x=0$ and $x\approx1$. The critical value of the parameter $r$ for the saddle-node bifurcation to occur lies at $r=\sqrt{\frac{4b^3}{27a}}\approx 0.19$. 
\section{Hastening tipping cascades with attractive HOI}
\subsection{Numerical Results }
\par We analyze the numerical results of our model on various undirected network topologies, viz., random or Erd\H{o}s--R\'enyi (ER), scale-free or Barab\'{a}si-Albert (BA), and the small world or Watts-Strogatz (WS) networks for $N=100$.  
\par 
The pairwise coupling strength $\epsilon_1$ is set at $0.05$. At such a low coupling strength, only the first tipping element reaches the alternative stable state as its corresponding control parameter $r$ is set at $0.2$. However, it is unable to set off a tipping cascade due to the low pairwise coupling strength between the elements. We calculate the fraction of tipped elements $F$, i.e., the number of tipped elements divided by the total number of elements $N$, for each network topology. 
\par
\begin{figure}
    \centering  \includegraphics[width=0.8\linewidth]{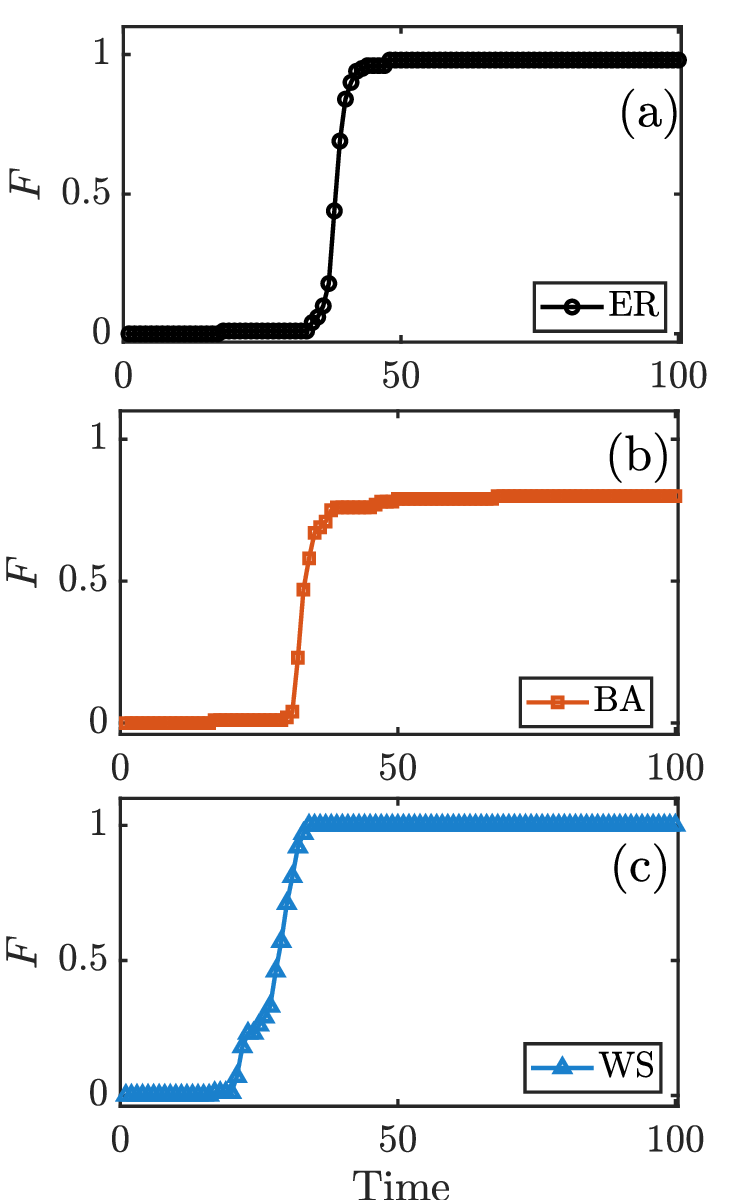}
    \caption{ The fraction of tipped elements $F$ with respect to the total number of elements $N$ is plotted with time. (a) The evolution of $F$ with time for the Erd\H{o}s--R\'enyi (ER) graph topology. We observe that nearly $100\%$ of the elements tip to an alternative state as illustrated by the black line with circular markers at $\epsilon_1=0.05$, $\epsilon_2=0.5$ and $\langle k_{ER}\rangle=6$. (b) For the scale-free or Barab\'{a}si-Albert (BA) network topology, nearly $60\%$ elements are found to have tipped, as displayed by the orange line with circular markers at $\epsilon_1=0.05$, $\epsilon_2=0.5$, and preferential attachment factor $m=2$. (c) The third subplot shows that almost the entire Watts-Strogatz (WS) network tips to the alternative state at $\epsilon_1=0.05$, $\epsilon_2=0.3$, $p_{WS}=0.2$, and each element is connected to $6$ nearest neighbors.  }
    \label{fig:2}
\end{figure}

 In Erd\H{o}s--R\'enyi or random networks, a link between elements $i$ and $j$ is added with a probability $p$. Hence, the average degree of the network can be calculated as $\langle k \rangle \approx p(N-1)$. For the ER network at $\epsilon_2=0.5$, $\epsilon_1=0.05$, and $\langle k_{ER}\rangle=6$, we observe a tipping cascade, illustrated by black lines with circular markers in Fig.~\ref{fig:2}(a), since the tipping elements coupled to the first element also reach the alternative stable state. Hence, it is observed that attractive higher-order interactions hasten tipping cascades, i.e., tipping of the entire network occurs due to the presence of higher-order interactions at low coupling strengths, at which tipping cascades do not occur with only pairwise interactions.  
\par Next, we generate a scale-free or a Barab\'{a}si-Albert network \cite{barabasi2009scale}. 
We observe that HOI impacts the occurrence of tipping cascades in scale-free networks, also with preferential attachment factor (i.e, new vertices attach preferentially to already well-connected elements) $m=2$ at $\epsilon_2=0.5$. We observe around $60\%$ of the elements in the network tip to the alternative state (Fig.~\ref{fig:2}(b)). 
\par A small-world or Watts-Strogatz network is also considered.
In Fig.~\ref{fig:2}(c), we observe that a tipping cascade occurs with all of the elements tipped for this network topology also at $\epsilon_1=0.05$, and $\epsilon_2=0.3$ with $p_{WS}=0.2$ with each element connected to $6$ nearest neighbors. Hence, in all three cases,  we observe that HOIs facilitate the occurrence of tipping cascades in networks at low coupling strengths where such cascades would be otherwise absent with pairwise interactions.  
\par
\begin{figure}
    \centering    
    \includegraphics[width=2.5in]{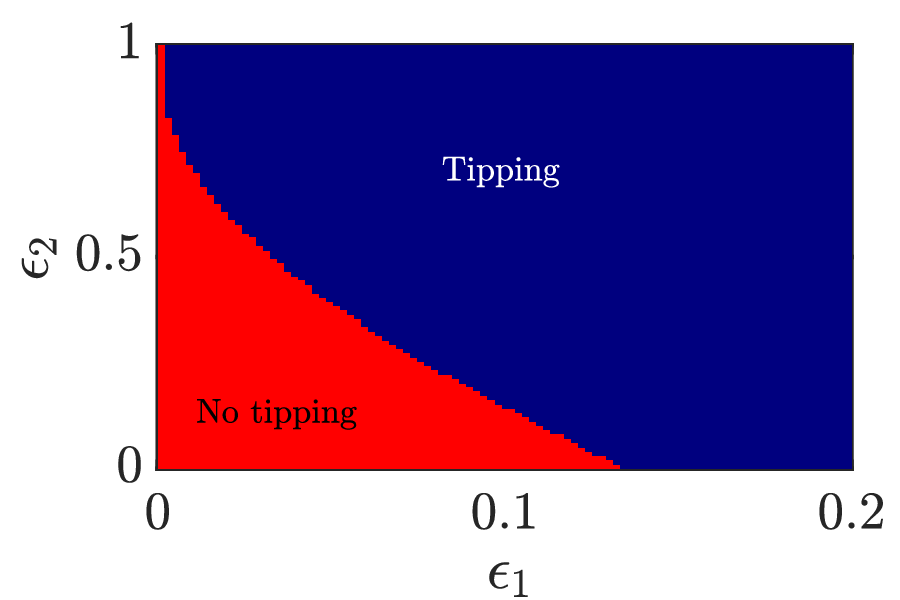}
    \caption{The parameter regime of $\epsilon_1-\epsilon_2$ is shown. The red regime shows the region where there is no tipping cascade, i.e., only the first element has tipped, and the rest are undisturbed. The dark blue regime displays the parameter region where the system undergoes a tipping cascade.  For each parameter pair, results are averaged over 100 random-network realizations; we label a point blue if a cascade occurs in at least $80\%$ of runs and red otherwise. }
    \label{fig:3}
\end{figure}

 Next, we investigate the interplay of both $1$-simplex and $2$-simplex interactions. Fig.~\ref{fig:3} shows the parameter regime for $\epsilon_1-\epsilon_2$. The parameter space has been computed for $100$ realizations of ER networks with an average degree of $4$. If at least $4$ elements of the system have tipped in $80\%$ of the runs for a parameter pair $(\epsilon_1,\epsilon_2)$, we label the point as ``tipping".  In Fig.~\ref{fig:3}, the red region represents the regime at which there is no tipping cascade, while the blue region shows the parameter space where tipping cascades take place. We observe that at $\epsilon_2=0$, the system tips at the critical coupling strength required for the system to undergo a tipping cascade is around $0.12$. It is also observed that as $\epsilon_2$ increases, the pairwise coupling strength required to facilitate a tipping cascade decreases. Thus, the parameter regime also reinforces our claim that HOIs encourage the occurrence of tipping cascades.
\subsection{Bifurcation analysis}
Next, we investigate the route of the tipping cascade with and without attractive higher-order interactions. To facilitate this, we explore the various bifurcations occurring in the system using the continuation software MATCONT \cite{dhooge2003matcont}. 
\par
\begin{figure*}[htp]
    \centering
    \includegraphics[width=0.9\linewidth]{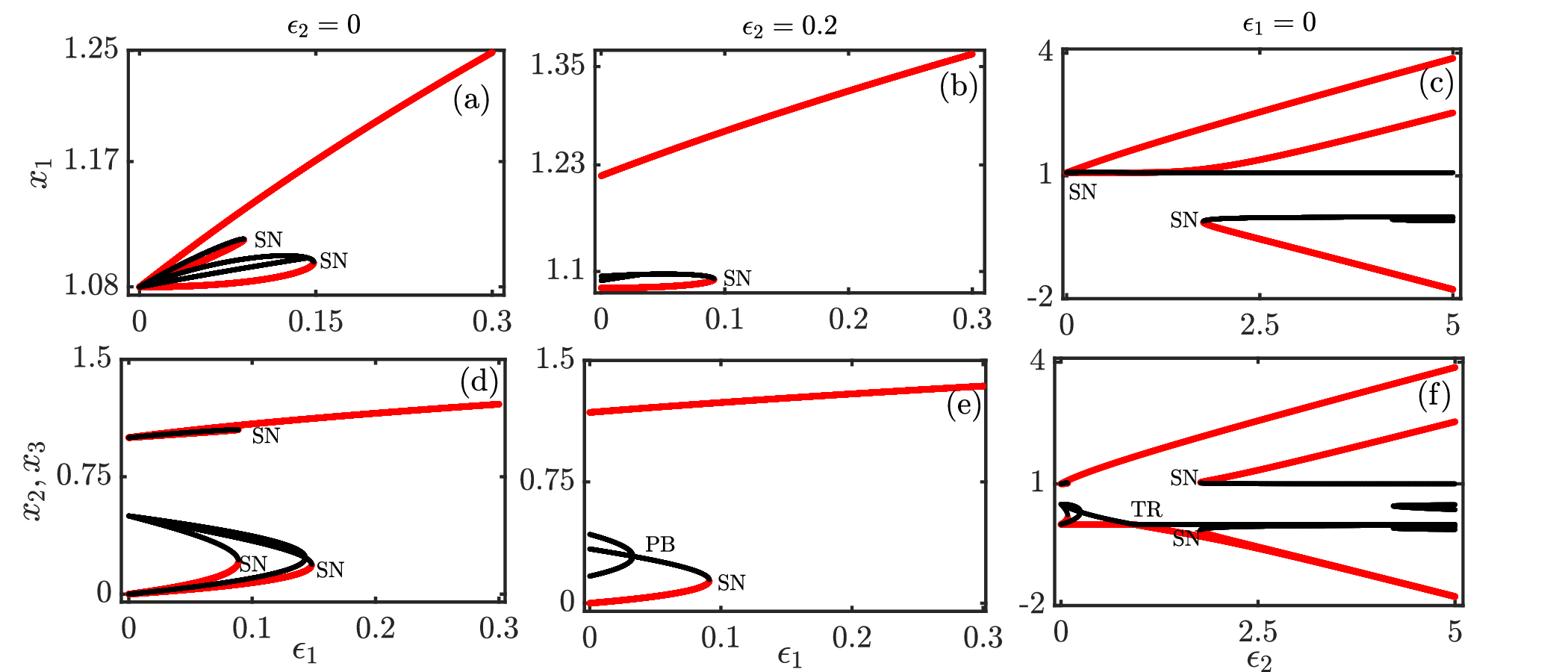}
    \caption{The bifurcation diagrams of the model with $N=3$ are presented here. The red lines represent the stable fixed points, whereas the black lines refer to the unstable ones.  The upper panel shows the bifurcations of the $x_1$ element, while the lower panel shows those of the identical $x_2$ and $x_3$ elements.  (a) We observe that the $x_1$ element undergoes saddle-node bifurcations (SN), and the three alternative states of the system collapse to a single one. (b) It is observed that a saddle-node bifurcation occurs at $\epsilon_1=0.1$ for a fixed higher-order coupling strength of $\epsilon_2=0.2$, and one of the alternative states gets destroyed. (c) When the pairwise coupling strength $\epsilon_1$ is absent, we find that saddle-node bifurcations occur with respect to $\epsilon_2$. (d) The elements $x_2$,$x_3$ initially are in the stable zero state, which gets destroyed by saddle-node bifurcations with $\epsilon_2=0$. The alternative state, which is around $1$, is also shown. (e) shows the bifurcations associated with $\epsilon_2=0.2$. We find that the saddle-node bifurcation occurs for $\epsilon_1 \approx 0.1$ due to HOI. (f) The system's stable zero state at $\epsilon_1=0$ undergoes a transcritical bifurcation (TR) and becomes unstable at around $\epsilon_2\approx 1$.}
    \label{fig:4}
\end{figure*}

 We consider the model with interactions between $N=3$ globally coupled elements only. As defined above, the control parameter of the first element is considered as $0.2$, while that for the rest of the elements is zero. Fig.~\ref{fig:4} depicts the various bifurcations occurring in all three elements of the system, i.e., $x_1$, $x_2$, and $x_3$. The first column shows the bifurcations without HOI, i.e., $\epsilon_2=0$,  the second column displays the bifurcations of the system with HOI at $\epsilon_2=0.2$. The third column shows the bifurcations with only HOI, i.e., no pairwise interactions are present in the system ($\epsilon_1=0$). 
\par In Fig.~\ref{fig:4}(a), it is observed that the $x_1$ element of the system, whose control parameter $r$ resides at $0.2$, resides in the alternative state because of the parameterization of $r$. The system has three alternative stable states to choose from, but two of them disappear at around $0.15$ through saddle-node bifurcations, leaving only one alternative stable state. In Fig.~\ref{fig:4}(d), the bifurcation diagrams of both the identical elements $x_2$ and $x_3$ are displayed. It is observed that the system's zero stable states are destroyed by consecutive saddle-node bifurcations, while another saddle-node bifurcation destroys an alternative state. Hence, the system has no choice but to occupy the remaining alternative states. The coupled elements also tip, and tipping cascades occur in the system due to the emergence of saddle-node bifurcations that arise as a consequence of the increase in pairwise coupling strength and the first element $x_1$. \par Figs.~\ref{fig:4}(b) and (e) display the bifurcations occurring in the system with both pairwise and higher-order interactions. We observe that due to the higher-order interaction coupling strength at $\epsilon_2=0.2$, one of the alternative stable states of $x_1$ disappears through a saddle-node bifurcation and the system is left with only one alternative stable state. In the elements $x_2$ and $x_3$, we observe that due to the presence of higher-order interactions at $\epsilon_2=0.2$, the critical coupling strength for the destruction of the zero stable state reduces to below $0.1$. Hence, we can say that higher-order interactions encourage tipping cascades to occur at lower coupling strengths than with only pairwise interactions. \par We also explore the bifurcations with only HOIs, i.e., when pairwise interactions are absent. We observe that saddle-node bifurcations occur both in the positive and negative $x-y$ plane. As with the other cases, alternative stable states are present for the $x_1$ node from the onset of the HOIs. However, for the elements $x_2$, $x_3$, the dynamics are a little different. We observe that the zero state remains stable till $\epsilon_2\approx1$. Then it is destroyed by a transcritical bifurcation denoted by $TR$, and the zero state becomes unstable. Hence, the system has no option but to occupy the alternative stable states. Thus, attractive HOIs not only hasten tipping cascades, but when they act alone without the influence of pairwise interactions, the route to the tipping cascade is also altered as a transcritical bifurcation occurs in this case instead of the usual saddle-node one.  
\subsection{Real world social network}
\begin{figure*}
    \centering
    \includegraphics[width=0.8\linewidth]{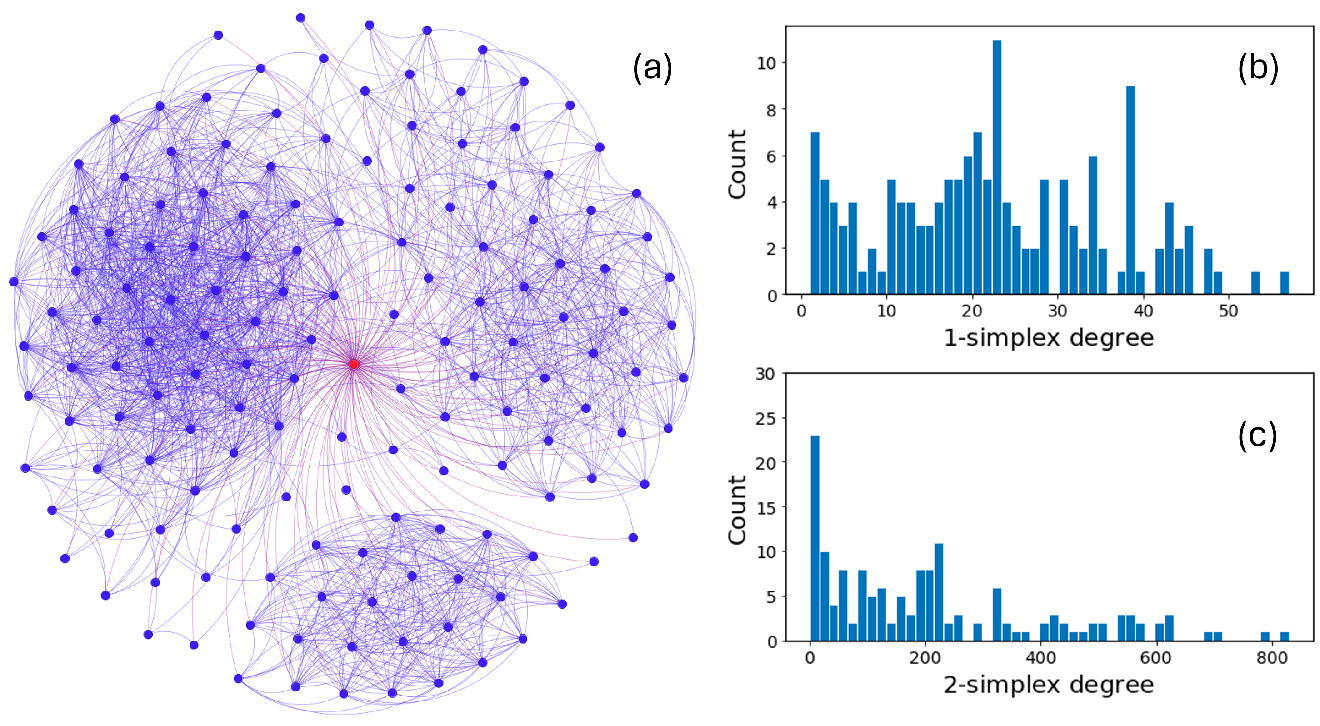}
    \caption{ (a) The social network of one anonymous Facebook user (dataset-$1$) has been presented here. The circular markers represent the elements, and the lines refer to the edges connecting the elements (nodes). The node highlighted in red corresponds to the ego node, representing the user whose social network is illustrated. This element is connected to all of the other elements in the network while the blue circular markers represent the ``friends" or ``alters" of the ego element. (b) The degree distribution of $1$-simplices or links is presented here. (c) The degree distribution of $2$-simplices or triangles is shown. }
    \label{fig:5}
\end{figure*}

Next, we investigate whether the phenomenon of tipping cascades in networks with attractive HOIs can occur in real-world networks as well. We chose social network datasets for this exercise, since HOIs are natural in social networks, especially social networking sites, where groups of people can interact simultaneously. The data used to build this network has been obtained from the website SNAP (Stanford Network Analysis Project) \cite{snapnets,leskovec2012learning}. We have used the freely available networking data of one anonymous Facebook user to construct the social network of the said user. That is to say, we build the network of people a certain user is connected to. The user is referred to as the ``ego" and the user's friends are referred to as ``alters" \cite{leskovec2012learning}. Fig.~\ref{fig:5}(a) displays the structure of the network. The blue elements are the alters or the friends of the user, and the ego element is colored in red. The blue lines refer to the edges or links from the neighbors, while the red lines represent the edges from the ego element. The properties of the network have been quantified in Fig.~\ref{fig:5}(b), where the degree distribution of the links or $1$-simplex has been displayed in the form of a histogram. For the higher-order interactions, we construct triangles or $2$-simplices from the pairwise interactions. We also quantify the density of $2$-simplices formed in Fig.~\ref{fig:5}(c). 
\par
\begin{figure}
    \centering
    \includegraphics[width=0.8\linewidth]{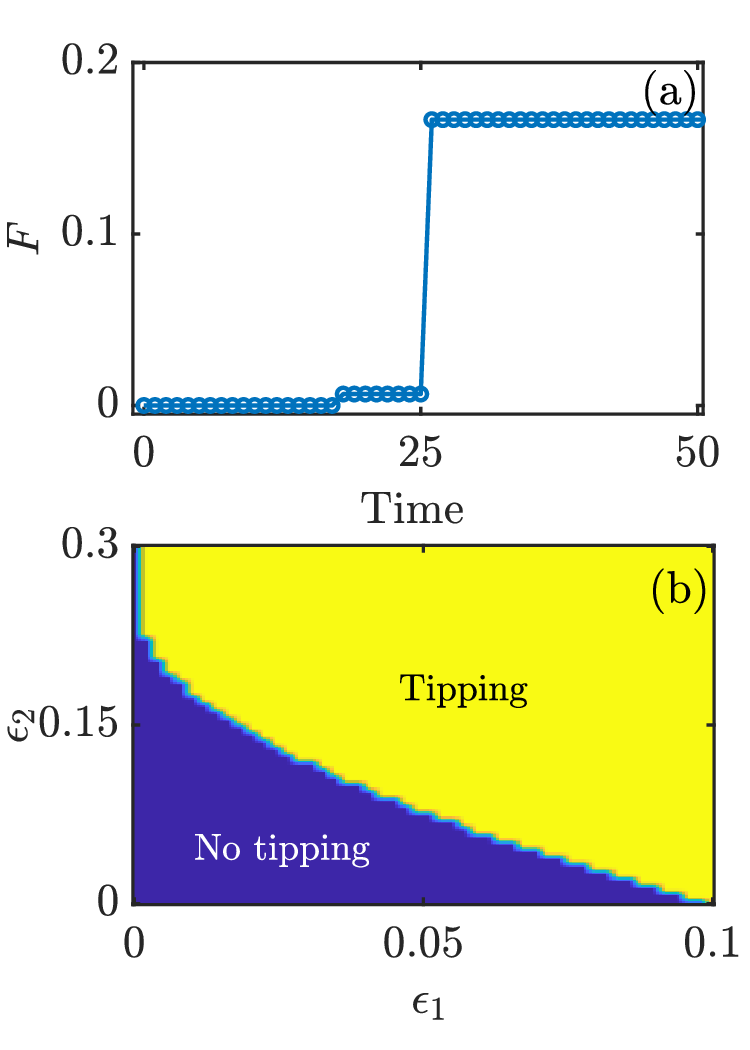}
    \caption{(a) The fraction of tipped elements for social network dataset-$1$ is shown. A small number of elements tip to the alternative state, whereas the rest remain undisturbed. (b) For the second social network dataset, i.e., dataset-$2$, we have presented the parameter regime of $\epsilon_1-\epsilon_2$. The blue regime shows the region where only the first element tips and no cascade occurs, while the yellow region refers to where the tipping cascades occur. If at least 4 elements tip to an alternative state, we label a parameter pair point as "tipping", and "no tipping" otherwise. }
    \label{fig:6}
\end{figure}

 Two datasets have been considered, and we label them dataset-$1$ and dataset-$2$. Networks have been constructed from both. In Fig.~\ref{fig:6}, we present the results for both datasets. 
\par We simulate our model over the networks constructed from the datasets. We first consider dataset-$1$. The pairwise coupling strength $\epsilon_1$ is taken as $0.002$. At this low value of $\epsilon_1$, only the first element tips because of $r$ at $0.2$, and the rest remain undisturbed. However, on the application of HOI coupling strength $\epsilon_2=0.05$, we observe that a tipping cascade occurs as observed in Fig.~\ref{fig:6}(a). Only a small number of nodes tip to the alternative state as a result of the pairwise and higher-order interactions. This outcome reflects real-world scenarios, where not all individuals are equally susceptible to influence. For instance, a user may share a sensational news item with their friends, but only a fraction of them may accept or be influenced by it, while others remain unaffected. 
For the second dataset, i.e., dataset-$2$, we have presented the parameter regime of $\epsilon_1-\epsilon_2$. We observe that tipping cascades occur at sufficiently high values of higher-order coupling strength, even for very low values of pairwise coupling strength. We label a point as "tipping" if at least $4$ elements tip to the alternative state and "no tipping" otherwise. Hence, it is observed that attractive HOIs promote the occurrence of tipping cascades in real-world network topologies as well. 
  
\section{Mitigating tipping cascades with repulsive HOI}
\par Further, we examine the influence of repulsive HOI on tipping cascades on complex networks. A real-world example of repulsive HOI might be in group decision-making, where the presence of a third person dampens the consensus that would emerge from a pair. For example, two people may form the same opinion, and the third may strongly disagree. We find that such repulsive interactions play a key role in suppressing cascade onset. \par
\begin{figure}
    \centering
    \includegraphics[width=0.8\linewidth]{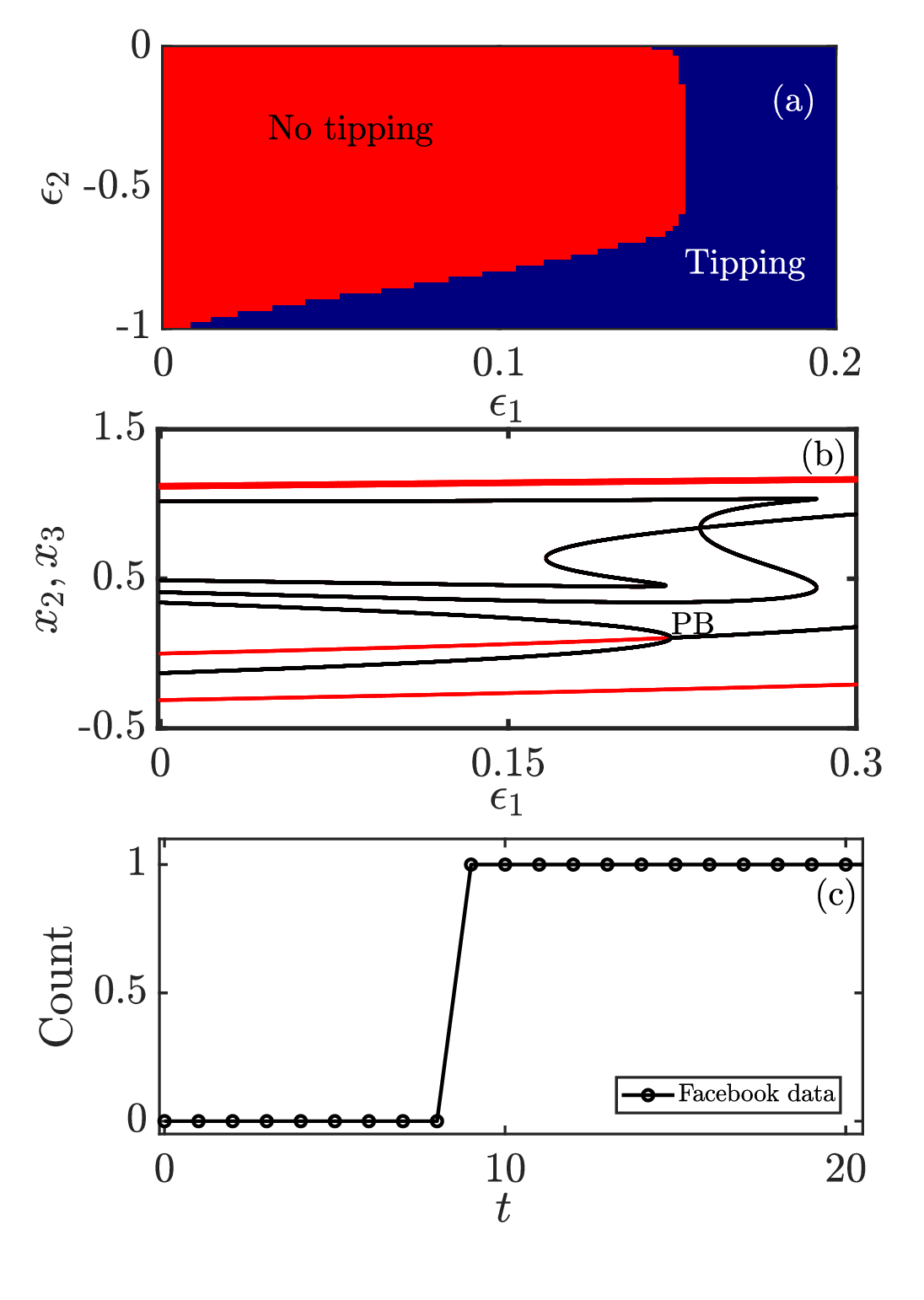}
    \caption{(a) Parameter regime of $\epsilon_1-\epsilon_2$ simulated over $100$ realizations of ER networks is presented. Here, $N=100$ and $\langle k \rangle =4$. We observe the onset of tipping cascades shift towards greater values of pairwise coupling strengths due to the repulsive HOI. The red regime presents the parameter region where no tipping cascade takes place, whereas the blue regime presents the parameter region where tipping cascades take place. (b) Bifurcation diagram with $N=3$ at $\epsilon_2=-0.5$ is displayed. The red lines correspond to the stable solutions, while the black ones refer to the unstable solutions. We observe that a supercritical pitchfork bifurcation, denoted by \textit{PB}, occurs, and the trivial state loses stability and the system occupies the alternative stable state around $1$. (c) The number of elements that tip on a network constructed from a social network dataset (dataset-$1$)  is presented. We observe that at $\epsilon_1=0.08$ and repulsive HOI $\epsilon_2=-0.05$, we show that only the first element tips and the rest of the elements are undisturbed. }
    \label{fig:7}
\end{figure}

  Fig.~\ref{fig:7} showcases the effect of such interactions on tipping cascades. Fig.~\ref{fig:7}(a) maps the parameter space of $\epsilon_1-\epsilon_2$. We have simulated the model over $100$ realizations of the ER network with $N=100$ and an average degree of $\langle k \rangle =4$. It is observed that in the presence of repulsive HOI, a higher value of pairwise coupling strength $\epsilon_1$ is required for a tipping cascade to occur. Consequently, the region of parameter space where no cascades occur expands, demonstrating that repulsive HOIs act to mitigate tipping cascades at parameters where they would have occurred with only pairwise interactions. 
 We also investigate the route to tipping cascades in the presence of repulsive HOI. For this, we consider an all-to-all coupled network of $N=3$. Fig.~\ref{fig:7}(b) shows the bifurcation diagram with respect to $\epsilon_1$ with a fixed repulsive HOI coupling strength of $\epsilon_2=-0.5$. We find that in the face of repulsive HOI, the route to tipping of the $x_2$ and $x_3$ elements changes from a saddle-node bifurcation to a supercritical pitchfork one. The stable zero state loses stability at around $0.2$ through a reverse supercritical pitchfork bifurcation and becomes unstable. Hence, the system has no choice but to occupy the alternative stable state at around $1$. Thus, repulsive HOI affects the dynamics of the system so profoundly that not only are tipping cascades mitigated, but the route to a tipping cascade is also altered.  The effect of repulsive HOI has also been explored in real social networks. Fig.~\ref{fig:7}(c) displays the count of elements that tips at $\epsilon_1=0.08$ and $\epsilon_2=-0.05$. When $\epsilon_2=0$, we observe a tipping cascade at this pairwise coupling strength. However, when the repulsive higher-order coupling strength is applied, it is observed that only the first element tips, and the tipping cascade is averted. Hence, repulsive HOIs play a decisive role in suppressing tipping cascades on complex networks. 

\section{Conclusion}

An element is said to have tipped if it occupies an alternate stable state from one stable state. A tipping cascade occurs when tipping elements tip due to their coupling with a tipped element. In this article, we have investigated the influence of higher-order interactions on the occurrence of cascades. Tipping cascades occur in different systems in nature, such as chains of lakes or power grids, and others. Such cascading effects have been explored in multiple systems, e.g., climate \cite{wunderling2024climate} and synthetic networks, including real-world networks obtained from the data of the atmospheric flow of moisture between different Amazonian forest cells \cite{kronke2020dynamics}. 
\par In our investigations, we have considered the cusp bifurcation model. The control parameter of one element has been set to $0.2$, while the control parameter $r$ of the rest of the elements is set to zero. It is observed that tipping cascades occur due to the interactions between the elements in a complex network. In this article, we have considered both pairwise and higher-order interactions. We have found that higher-order interactions promote the occurrence of tipping cascades, i.e., such cascades arise at lower coupling strengths than in systems with only pairwise interactions. \par We have also explored the bifurcations associated with the tipping cascades. We find that the route to tipping is through a saddle-node bifurcation with respect to the pairwise coupling strength $\epsilon_1$. When we turn off the pairwise coupling, i.e., $\epsilon_1=0$ and observe the bifurcations with only higher-order interactions, we find that the route to tipping completely changes. A transcritical bifurcation occurs, and the trivial state loses stability, leaving the system to occupy any of the alternative stable states. Thus, higher-order interactions can alter the pathway through which the system transitions to an alternative stable state as well. 
\par We have also explored the system on real-world social networks, the data of which have been obtained from SNAP (Stanford Network Analysis Project) \cite{snapnets}. We have considered the social networks of two anonymous users from the social networking site Facebook. We observe that tipping cascades arise at lower coupling strengths on real social networks when both pairwise and higher-order interactions are present, compared to networks with only pairwise interactions. Hence, it can be said that HOIs hasten the occurrence of tipping cascades. 
\par We have also considered the model with repulsive higher-order interactions. In this case, we find that tipping cascades are mitigated, and the parameter space regime where tipping cascades do not occur expands in the presence of the repulsive HOI. In fact, we observe that cascades are suppressed under repulsive HOI in regimes where they would otherwise emerge in the absence of such interactions. We also find that the route of such tipping cascades is changed from a saddle node to a supercritical pitchfork. It is also shown that tipping cascades are mitigated at parameters where they would otherwise emerge without such interactions in a real-world social network. In real-world systems, repulsive higher-order interactions can arise in various ways to control or mitigate tipping cascades. In the context of social networks, they may manifest as negative opinions that counteract consensus formation. In biological systems, they can represent inhibitory influences exerted by a species that weaken the collective effect of other species. Such repulsive interactions can act as a control mechanism for tipping cascades and especially help in mitigating such cascades, since tipping cascades can be undesirable in many scenarios. 

Thus, in conclusion, we find that attractive higher-order interactions promote the occurrence of tipping cascades while repulsive higher-order interactions act to mitigate such cascades in complex networks. The route to tipping cascades is also altered in the presence of only attractive higher-order interactions from saddle-node to transcritical. For repulsive interactions, we find that the route is changed to a supercritical pitchfork bifurcation. These results are also observed in real-world social networks. This work can be further extended by considering multi-layer topologies or ecological systems, which will be reported in further studies. We hope our results will be useful to theoretical ecologists, climate scientists, and policy-makers who work with tipping cascades in complex networks. 
\section{Acknowledgements}
RG acknowledges financial support from the University Grants Commission, New Delhi (Grant number: UGCES-22-GE-WES-F-SJSGC-1732). MDS acknowledges financial
support from SERB, Department of Science and 
Technology (DST), India (Grant No. CRG/2021/003301).
\bibliography{main} 
\end{document}